\RequirePackage[loading]{tracefnt}
\documentclass[fleqn,usenatbib,useAMS]{mnras}

\usepackage[T1]{fontenc}
\usepackage{ae,aecompl}
\usepackage{newtxtext,newtxmath}

\usepackage{graphicx}	
\usepackage{amsmath}	
\usepackage{amssymb}	
\usepackage{multicol}        
\usepackage{bm}		
\usepackage{pdflscape}	

\usepackage{color}

\usepackage{etoolbox}
\makeatletter
\patchcmd\@combinedblfloats{\box\@outputbox}{\unvbox\@outputbox}{}{\errmessage{\noexpand patch failed}}
\makeatother

\title[Molecular clumps and mixing]{Free-floating molecular clumps and gas mixing: hydrodynamic aftermaths of the intracluster--interstellar medium interaction}
\author[Ruggiero et al.]{Rafael Ruggiero$^{1,2}$\thanks{E-mail: rafael.ruggiero at usp.br}, Romain Teyssier$^2$, Gastao B. Lima Neto$^1$ and Valentin Perret$^2$
\\
$^{1}$Instituto de Astronomia, Geof\'isica e Ci\^encias Atmosf\'ericas, Universidade de S\~ao Paulo, R. do Mat\~ao 1226, 05508-090 S\~ao Paulo, Brazil\\
$^{2}$Institute for Computational Science, Centre for Theoretical Astrophysics and Cosmology, Universit\"at Z\"urich, CH-8057, Z\"urich, Switzerland\\
}
\date{Accepted 2017 March 23. Received 2017 March 21; in original form 2016 December 12} 
\pubyear{2017}

\begin{document}
\label{firstpage}
\pagerange{\pageref{firstpage}--\pageref{lastpage}}
	\maketitle

\begin{abstract}

The interaction of gas-rich galaxies with the intra-cluster medium (ICM) of galaxy clusters has a remarkable impact on their evolution, mainly due to the gas loss associated with this process. In this work, we use an idealised, high-resolution simulation of a Virgo-like cluster, run with RAMSES and with dynamics reproducing that of a zoom cosmological simulation, to investigate the interaction of infalling galaxies with the ICM. We find that the tails of ram pressure stripped galaxies give rise to a population of up to more than a hundred clumps of molecular gas lurking in the cluster. The number count of those clumps varies a lot over time -- they are preferably generated when a large galaxy crosses the cluster (M$_{200c} > 10^{12}$ M$_\odot$), and their lifetime ($\lesssim 300$ Myr) is small compared to the age of the cluster. We compute the intracluster luminosity associated with the star formation which takes place within those clumps, finding that the stars formed in all of the galaxy tails combined amount to an irrelevant contribution to the intracluster light. Surprisingly, we also find in our simulation that the ICM gas significantly changes the composition of the gaseous disks of the galaxies: after crossing the cluster once, typically 20\% of the cold gas still in those disks comes from the ICM.

\end{abstract}


\section{Introduction}

The environment of a galaxy is an important driver for its evolution across a range of scales. One recurrent scenario is that of a small satellite orbiting a large halo and being affected by it, which can be a dwarf galaxy orbiting a massive one or a galaxy orbiting a galaxy cluster \citep[see][]{1972ApJ...176....1G,1998ARA&A..36..435M}. These interactions give rise to an universal dependence of galaxy properties on local density \citep{2004MNRAS.353..713K,2013MNRAS.432..336W}, suggesting that, in a model of galaxy evolution, the environment must be taken into account in order to explain the observed properties of galaxies in the universe.

In the case of galaxy clusters, orbiting galaxies suffer the influence of the diffuse and hot gaseous medium of these structures, known as the intracluster medium (ICM). As described in \citet{1972ApJ...176....1G}, the ram pressure exerted by the ICM can remove gas from the galaxies, in a process called ram pressure stripping. This process has been extensively reported in observations, and is often associated with remarkable ``jellyfish'' morphologies, in which the galaxies feature a tail of stars and stripped gas organised in filaments and clumps \citep{2014ApJ...781L..40E, 2016MNRAS.455.2994M, 2016AJ....151...78P}. For a gas-rich disk galaxy, the ram pressure strips first the less gravitationally bound outskirts of its gaseous disk, causing it to feature a truncated H$_{\mathrm{I}}$ disk after a partial stripping event, such as has been observed in several Virgo \citep{1990AJ....100..604C,2004AJ....127.3361K} and Coma \citep{2000AJ....119..580B} cluster galaxies. The remaining gas, on the other hand, can in some cases be shock compressed by the pressure, giving rise to an increase in the star formation rate and luminosity of the galaxy \citep{2014ApJ...781L..40E,2017A&A...605A.127P}.

The tails of ram pressure stripped galaxies have been observed in several wavelengths, for instance in H$_\mathrm{I}$ for Virgo cluster galaxies \citep{1990AJ....100..604C,2007ApJ...659L.115C}, in H$_\alpha$ for a sample of Coma cluster galaxies \citep{2010MNRAS.408.1417S} and in X-rays \citep[e.g.][]{2010ApJ...708..946S}. It has been suggested that the tails can remain partly in a molecular, star-forming state even after the galaxy is already several tens of kpc away from the stripped region, although the star formation within this stripped gas is presumably inefficient \citep{2015A&A...582A...6V,2017ApJ...839..114J}. A prominent example is the large ($110 \times 25$ kpc) cloud of H$_\mathrm{I}$ gas which has been observed near the centre of the Virgo cluster \citep{2005A&A...437L..19O}, and which is likely to be the result of the ram pressure stripping of the nearby NGC 4388 galaxy. Another possible source of intracluster molecular gas are cooling-flows, as shown by radio observations of the central region of cool-core clusters \citep[e.g.][]{2001MNRAS.328..762E,2003A&A...412..657S}. Tail-like features can also be associated with tidal interactions between two or more galaxies in a cluster, as has been observed in Virgo \citep{2012A&A...544A..99W} and Shapley \citep{2016MNRAS.460.3345M}, although these features tend to be more local, not stretching over large distances like ram pressure tails, and to not be directly associated with a clumpy morphology.

Recently lone clumps of cold gas have been reported in the Virgo cluster, without obvious association with any galaxy
\mbox{\citep{2010ApJ...725.2333K,2017ApJ...843..134S}}, and also in other massive clusters \citep{2017ApJ...846L...8M}.
These clumps are low in mass (typically M$_{\mathrm{HI}} < 10^9$ M$_\odot$), and either do not have any optical counterpart or feature an exclusively young stellar population, indicating that they are not simply dwarf galaxies which have been accreted into the cluster, but instead are byproducts of relatively recent stripping events. They are also preferentially found in the outskirts of their host galaxy clusters. \citet{2017ApJ...843..134S} refer to the objects they report as Ultra Compact High Velocity Clouds (UCHVCs), a term most commonly used for similar systems in the local group \citep[e.g.][]{2013ApJ...768...77A,2015ApJ...806...95S}. Similar objects have also been reported in the intra-group medium of other galaxy groups \citep{2008AJ....135..319D,2009A&A...507..723T}, which could possibly have an analogous origin to the clouds observed in much more massive systems like the Virgo cluster.

Several numerical works have been performed to characterise the ram pressure stripping process. Regarding the ram pressure tails, \citet{2008MNRAS.388..465R} have studied in detail their dynamics, showing that they are significantly turbulent and feature a flaring spatial distribution. \citet{2011ApJ...731...98T} have shown that these tails are not always expected to be bright in X-rays, since that depends on the local pressure -- higher densities favouring the appearance of X-ray emission due to a favourable mixing between the stripped gas and the ICM. The local pressure has also been shown to affect the star formation rate in the tail \citep{2012MNRAS.422.1609T}.

The novelty of the work we present here is to use a realistic setup, including an entire spherical galaxy cluster \citep[as in][]{2008MNRAS.388..465R,2017MNRAS.468.4107R} and infalling galaxies extracted from a cosmological simulation, to characterise in detail the clumps of gas left behind by these galaxies, including both their internal and dynamic properties. We focus not on the clumps which have just been ejected from the galaxy, but on the ones which survive for long enough to distance themselves from it. We additionally use our data to predict the level of contamination of ICM gas within their gaseous disks as they interact with the cluster, and the contribution of the star formation in the clumps to the intracluster light.

We structure this paper in the following manner. In Section \ref{sec:simulations} we describe the initial conditions and numerical methods of our simulations. In Section \ref{sec:results} we show our results, which are divided into those regarding the clumps, those regarding the galaxies and a brief analysis of the intracluster light; and in Section \ref{sec:discussion} we discuss the results and summarise.

\section{Simulations} \label{sec:simulations}

This work involves two simulations. The first is a dark matter only, zoom cosmological simulation of a Virgo-like galaxy cluster, which is used as an input for the main simulation: an idealised, high resolution re-simulation of this same cluster including detailed gas physics. This second simulation features the infalling galaxies from the cosmological cluster, modelled as gas-rich disk galaxies, allowing us to study in detail the gaseous interactions between those galaxies and the ICM of the cluster. The density profile of the idealised cluster reproduces that of the cosmological one, and the infalling galaxies are the same both in halo mass, entry time and 3D entry location/velocity. We have additionally run a lower resolution version of the idealised simulation in order to assess its numerical convergence. All simulations were run with the Adaptive Mesh Refinement (AMR) code \textsc{RAMSES} \citep{2002A&A...385..337T}.

\subsection{Zoom cosmological simulation}

In our cosmological simulations, we assume a $\Lambda$CDM cosmology with $\Omega_m$ = 0.3, $\Omega_\Lambda$ = 0.7 and $H_0 = 70$ km s$^{-1}$ Mpc$^{-1}$ , consistent with Planck 2015 data \citep{2016A&A...594A..13P}. The initial conditions are generated with the code \textsc{MUSIC} \citep{2011MNRAS.415.2101H}, considering a cosmological box with side length of 70 Mpc $h^{-1}$ and only including dark matter particles. The simulations start at $z = 30$, and a second order Lagrangian perturbation theory is used to set the initial positions and velocities of these particles.

We first ran a series of periodic box cosmological simulations with $2^8\times 2^8 \times 2^8$ ($\sim1.6 \times 10^{7}$) particles each to search for a suitable cluster to zoom. The selection was based on the properties of the clusters at $z = 0$. The criteria were that the cluster had a similar mass to the Virgo cluster, for which the $M_{200c}$ (defined as the mass within the radius where the average density of the object is 200 times the critical density of the universe) has recently estimated to be $1.05\times10^{14}$ M$_\odot$ \citep{2017MNRAS.469.1476S}; that it did not have any major substructure, ensured by requiring that no sub-halo with $v_\mathrm{max} > 300$ km/s was present within the virial radius, so that it was as undisturbed and hence generic as possible; and that no large structure could be seen close to it, for the same reason. Our selected cluster fulfilled all of these criteria. 

We then generated the zoom initial conditions with \textsc{MUSIC} using as an input a convex hull containing all of the selected cluster's particles, as identified with the code \textsc{hast}\footnote{\url{https://bitbucket.org/vperret/hast}}. In the zoom simulation, two levels of refinement were added to the zoom region, and one level was removed from the base grid. Between two levels of refinement, a padding of 4 cells was added. The resulting mass resolution in the zoom region is $3.4 \times 10^{7}$ M$_\odot$. After the simulation, the resulting $M_{200c}$ of the zoom cluster at $z = 0$ was $1.03\times10^{14}$ M$_\odot$, very close to the Virgo cluster. Its virial radius is 1023 kpc, and its $R_{200c}$ is 1126 kpc.

For the purpose of the idealised re-simulation, we consider the last 5 Gyr of evolution of this zoom cluster, between redshifts $z = 0.5$ and $z = 0$. In the beginning of this interval, it has already reached 80\% of its final mass, and from then on it grows in mass gradually and without any major disturbance to its dynamical state. We identify all halos in the relevant snapshots using the \textsc{ROCKSTAR} halo finder \citep{2013ApJ...762..109B}, and select the ones with $M_{200c}$ above $10^{11}$ M$_\odot$ which cross the virial radius of the cluster for the first time in this interval. These halos will represent the infalling galaxies which will be used in our idealised re-simulation. They are shown in Table \ref{tab:galaxies}, along with other structural parameters that will be described in Section \ref{sec:idealised}.

\begin{table}
 \caption{The infalling galaxies extracted from the zoom cosmological simulation. The first four columns were extracted directly from the simulation, and the last two were computed having as an input the $M_{200c}$. Here $v_0$, $t_0$ and $b_0$ are the velocity, time and impact parameter at entry respectively.}
 \label{tab:galaxies}
 \resizebox{\columnwidth}{!}{%
 \begin{tabular}{llllll}
  \hline
  $M_{200c}$ (M$_\odot$) & $v_0$ (km/s) & $t_0$ (Myr) & $b_0$ (kpc) &  $M_\star$ (M$_\odot$) & $R_d$ (kpc) \\
  \hline
1.8 $\times 10^{12}$ & 1026 & 0 & 507 & 3.9 $\times 10^{10}$ & 3.48\\
1.1 $\times 10^{12}$ & 980 & 2330 & 248 & 2.6 $\times 10^{10}$ & 2.92\\
3.8 $\times 10^{11}$ & 917 & 1920 & 424 & 7.1 $\times 10^{9}$ & 1.65\\
3.1 $\times 10^{11}$ & 524 & 490 & 351 & 4.7 $\times 10^{9}$ & 1.38\\
2.9 $\times 10^{11}$ & 928 & 1760 & 720 & 3.8 $\times 10^{9}$ & 1.25\\
2.6 $\times 10^{11}$ & 1102 & 90 & 483 & 3.0 $\times 10^{9}$ & 1.13\\
2.0 $\times 10^{11}$ & 991 & 430 & 764 & 1.6 $\times 10^{9}$ & 0.86\\
1.7 $\times 10^{11}$ & 979 & 490 & 114 & 1.3 $\times 10^{9}$ & 0.77\\
1.7 $\times 10^{11}$ & 1032 & 2990 & 721 & 1.2 $\times 10^{9}$ & 0.76\\
1.6 $\times 10^{11}$ & 935 & 490 & 600 & 1.1 $\times 10^{9}$ & 0.71\\
1.4 $\times 10^{11}$ & 1066 & 2990 & 639 & 7.9 $\times 10^{8}$ & 0.63\\
1.3 $\times 10^{11}$ & 948 & 2540 & 678 & 7.1 $\times 10^{8}$ & 0.60\\
1.2 $\times 10^{11}$ & 993 & 1450 & 738 & 6.2 $\times 10^{8}$ & 0.56\\
1.1 $\times 10^{11}$ & 997 & 4070 & 464 & 4.7 $\times 10^{8}$ & 0.50\\
\hline
 \end{tabular}}
\end{table}

\subsection{Idealised simulation} \label{sec:idealised}

The initial conditions for our idealised simulation were generated with the code \textsc{dice}\footnote{\url{https://bitbucket.org/vperret/dice}}. The output of this code contains particle data, which is read into RAMSES using the \textsc{dice} patch\footnote{\url{https://bitbucket.org/vperret/dice/wiki/RAMSES\%20simulation}}. They consist of a spherical galaxy cluster in equilibrium, representing the zoom cluster, and the infalling galaxies described in the previous section. In order to ensure that the galaxies will enter the virial radius of the cluster at the same time and location as in the zoom simulation, we integrate their orbits back in time considering the potential of the cluster and their 3D position and velocity at entry, and place them in the initial conditions in the appropriate location 5 Gyr before $z = 0$.

The galaxies are made of a dark matter halo following a Hernquist density profile \citep{1990ApJ...356..359H}
\begin{equation} \label{eq:hernquist}
\rho(r) = \frac{M}{2\pi} \frac{a}{r} \frac{1}{(r+a)^3},
\end{equation}
with scale length $a$ fit to the cosmological halos; a stellar disk following an exponential profile
\begin{equation}
\rho(R, z) = \frac{M}{4\pi R_d^2 z_0} e^{-R/R_d} e^{-z/z_0};
\end{equation}
and a gaseous disk also following this profile, but with a different scale height, as we will describe later. The 3D orientation of the disk of each galaxy is randomised. A gaseous halo is not included; the evolution of gaseous halos of galaxies falling into clusters has been characterised in detail in \citet{2008MNRAS.383..593M}, but for simplicity we choose to assume that those halos are of secondary relevance to the evolution of the gaseous disks of infalling galaxies after they have entered the cluster, which is an assumption consistent with the recent simulations presented in \citet{2016A&A...591A..51S}, according to which the halo gas is expected to be nearly completely stripped within a very short timescale ($\sim 50$ -- $150$ Myr). The masses of the dark matter particles, the star particles and the gas particles are $5 \times 10^{5}$ M$_\odot$, $2.5 \times 10^{4}$ M$_\odot$ and $5 \times 10^{3}$ M$_\odot$, respectively. The gas particles are only used to initialise the simulation, as they are deleted after their mass is transferred to the simulation grid by the \textsc{dice} patch. Their mass is particularly small to ensure that their spatial coverage is high.

The stellar masses $M_\star$ assigned to the infalling halos extracted from the cosmological simulation are based on abundance matching data from \citet{2013ApJ...770...57B}, and the gas masses are fixed as $M_\mathrm{gas} = 0.2 M_\star$. The mass of each dark matter halo is set such that the $M_{200c}$ of the galaxy as a whole is the same of its counterpart in the zoom. To define the scale length $R_d$ of each disk, we use data from \citet{2010MNRAS.406.1595F}, in which the relation between stellar mass and disk scale length $R_d$ for a sample of non-interacting disk galaxies from the SDSS survey \citep{2000AJ....120.1579Y} was derived. In particular, we approximate the relation between $r$-band scale length and stellar mass by the following fit:
\begin{equation}
\log(R_d) = 0.4382 \log(M_\star) - 4.101.
\end{equation}
Once the stellar mass of the galaxy is known, we then use this relation to define the scale length of its disk, which is assumed to be the same for stars and gas. The vertical scales of the stellar and gaseous disks are defined as 10\% and 5\% of the radial scale length, respectively.

The galaxy cluster halo itself is replaced by an external potential representing its dark matter profile, along with a gaseous component modelled after the ICM of the Virgo cluster, considering the data presented in \citet{2011MNRAS.414.2101U}. The dark matter profile is represented by a Hernquist density profile (Equation \ref{eq:hernquist}), for which the best fit parameters were $M = 2.43 \times 10^{14}$ M$_\odot$ and $a = 231$ kpc. The observational ICM profile on the other hand is represented by a $\beta$-model \citep{1976A&A....49..137C}
\begin{equation} \label{eq:betaprofile}
\rho(r) = \rho_0 \left[1 + \left(\frac{r}{r_c}\right)^2\right]^{-\frac{3\beta}{2}}
\end{equation}
with the following fitted parameters: $\rho_0 = 1.1\times 10^{5}$ M$_\odot$ kpc$^{-3}$, $\beta = 0.45$ and $r_c = 38$ kpc. This reproduces well the outer slope of the observed profile ($r > 50$ kpc), while resulting in 10\% gas mass fraction within the virial radius relative to the cosmological cluster, which is a typical value for the cluster mass we are considering \citep{2013A&A...555A..66L}. The core radius $r_c$ of our fit is relatively small, but this is not important since different values would only affect galaxies galaxies falling with impact parameters close to zero, which do not exist in our simulation; also near the centre of the cluster, additional processes could be taking place which are beyond the scope of this paper (the presence of a BCG, AGN feedback, cooling-flows, etc). The mass of the ICM is subtracted from the mass of the Hernquist profile in such a way that the $M_{200c}$ of the cosmological cluster is conserved, while keeping the scale length of the profile unchanged. Additionally, we multiply the ICM density profile by the following function:
\begin{equation}
C(r) = \left[1+\exp\left(\frac{r-1103}{12.67}\right)\right]^{-1}.
\end{equation}
The purpose of this is to create a boundary for the cluster, so that the infalling galaxies only suffer its influence once they enter its virial radius. For $r < 1023$ kpc, $C(r) \approx 1$. As we will describe ahead, the ICM is resolved at a resolution of 31.25 kpc; $C(r)$ was designed such that for $r$ larger than the virial radius plus 5 of those cells (i.e. $r > 1023 + 5 \times 31.25$ kpc), $C(r) \approx 0$, thus creating a boundary that is well resolved by the resolution of the cluster. The number of particles used in the ICM initial conditions file is $10^7$, a high number for the sake of spatial coverage, analogously to the gas particles in the galaxies; and the temperatures of those initial ICM particles were calculated to ensure hydrostatic equilibrium. This results in a temperature of $\sim 9\times10^7$ K at the very central region of the cluster, which smoothly falls to $\sim 7 \times 10^6$ K at its virial radius.

The simulation is run in a box with side length of 8000 kpc. It follows, additionally to the regular gas quantities, a passive scalar, which is initially defined as 1 in the gas cells of the disks of the galaxies and 0 elsewhere. As the simulation runs, the scalar is advected with the gas, and we use it to restrict the refinement to the gas coming from the galaxies, ensured by checking if the value of the scalar in each cell is larger than $10^{-3}$. The ICM is thus left unrefined, being resolved only by a base grid with 8 levels of refinement, corresponding to a cell size of 31.25 kpc. The maximum level of refinement is 16, corresponding to a maximum resolution of 122 pc, and a padding of 5 cells is added between each level of refinement. A cell is refined if it contains more than 8 particles, which ensures that the collisionless components of the simulation (dark matter and stellar particles) are always resolved; or if its size is larger than the local Jeans length divided by 64, which is enough to force all the cold gas in the simulation (e.g. the gas in the disks of the galaxies) to be refined up to the maximum level of refinement. We do not employ a mass-based refinement criterion for the gas in the simulation, since that would overly resolve the tails of the galaxies, making the simulation intractable. This means that although the grid structure is made very dense within and around the cold gas in the tails, regions in an advanced state of mixing with the ICM, which are of secondary importance for our purposes, will have their internal structure resolved in less detail.

Radiative cooling is included but restricted to the cells associated with the galaxies, using the same passive scalar threshold as for the refinement. The purpose of this restriction is to prevent a cooling-flow from happening in the ICM. The cooling rate is calculated assuming a constant metallicity of 1 Z$_\odot$, and a cooling floor of 100 K is used to prevent an artificial runaway cooling of the gas. Star formation is modelled by a Schmidt law \citep{1959ApJ...129..243S}, which is triggered in gas cells with density above $n_\star$ = 0.1 H/cm$^3$, generating star particles with an efficiency per free fall time $\epsilon_\star$ = 0.01, and with a minimum particle mass of $5.6 \times 10^3$ M$_\odot$. We restrict star formation to within the virial radius of the cluster, so that the galaxies do not consume the gas in their initial conditions before entering the cluster. A simple sub-grid supernova feedback recipe is also included. The recipe is the same as the one described and used in \citet{2013MNRAS.429.3068T}, and it injects thermal energy into the simulation considering, in our case, that 20\% of the stellar mass formed explodes as supernovae, a fraction consistent with a Chabrier initial mass function \citep{2003PASP..115..763C}. Each supernova event releases $10^{51}$ erg of thermal energy, with an average supernova progenitor mass of 10 M$_\odot$, which combined with the supernova mass fraction results in a specific supernovae energy rate of $10^{49}$ erg per solar mass. A caveat of this feedback recipe is that it is inefficient in heating the cold gas around a stellar particle which undergoes a supernova event, since most of the injected energy is lost due to radiative cooling in a short timespan \citep[see e.g.][]{2009ApJ...695..292C}.

\section{Results} \label{sec:results}

In analysing both the clumps and the gas in the galaxies, we consider only the gas cells within those objects for which the density is above a self-shielding threshold. This threshold corresponds to the mean density of a multi-phase gaseous cloud above which the Hydrogen is expected to be mostly cold and in a molecular state, with the cloud self-shielding all its own ionizing radiation. Following \citet{2017arXiv170706993C}, we choose a typical value of $n_\mathrm{H} = 3 \times 10^{-3}$ cm$^{-3}$ (or $\rho = 5 \times 10^{-27}$ g cm$^{-3}$) for this threshold.

\subsection{Clumps}

The clumps in our simulation are found using the \textsc{PHEW} clump finder \citep{2015ComAC...2....5B}, which is currently built into \textsc{RAMSES}. We set it to find over-densities with a relevance threshold of 3 and with a saddle threshold of $1 \times 10^{-26}$ g cm$^{-3}$. The first value is relatively low so that no clump is lost; spurious clumps will be handled subsequently. The second value, which regulates the merging of nearby peaks, was empirically found to merge efficiently clumps which visually overlap with each other. 

For each snapshot, we proceed to discard the clumps within a radius of $10 R_d$ of any galaxy, since those nearby clumps are highly filamentary and entangled with one another, and we are interested here in those which are more isolated and relaxed. We also discard clumps without any self-shielded cells, since those are either spurious or in an advanced state of mixing with the ICM. For each remaining clump, we then assign a bounding sphere in the following manner. First we create a sphere centred in the clump position with radius equal to the smallest cell size in the simulation (122 pc). Then we gradually increase the radius of this sphere until only 50\% of the cells contained in it are self-shielded. This way, the sphere will encapsulate the whole boundary even of clumps which are not exactly spherical. After all bounding spheres are found for a snapshot, we then discard the clumps for which their spheres overlap with another sphere, thus ensuring that our results are not contaminated by interacting clumps. Those clumps are a tiny minority of the whole.

\begin{figure*}
 \includegraphics[width=1.63\columnwidth]{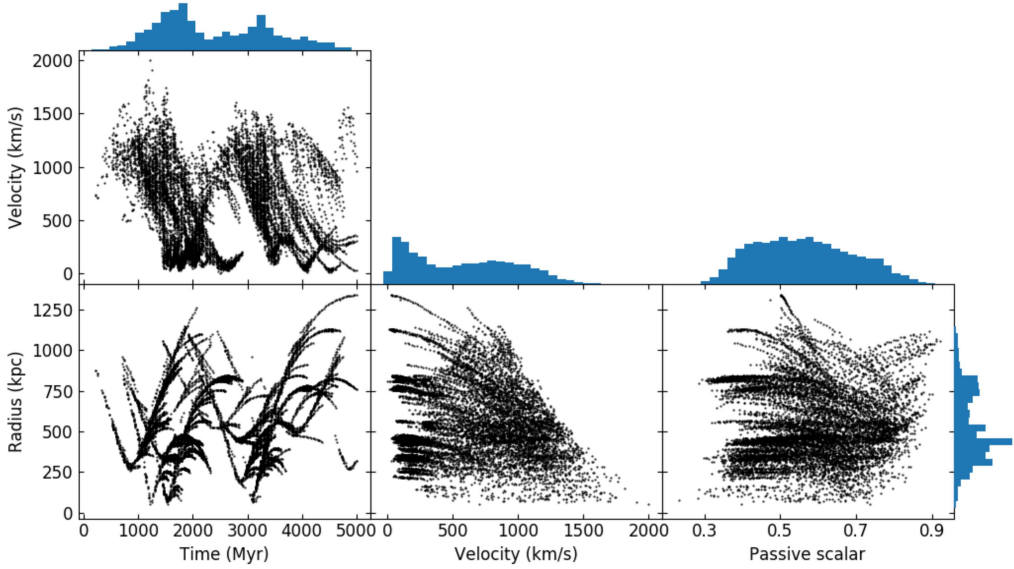}
 \caption{The dynamic evolution of the clumps of cold gas present in our idealised simulation. \emph{Passive scalar} refers to the mass-weighted average of the passive scalar in the cells of a clump.}
 \label{fig:clumpskinematics}
\end{figure*}

In Figure \ref{fig:clumpskinematics} we show the dynamic properties of the clumps. They can be found anywhere in the cluster, but they are more likely to be at $r < 500$ kpc, where they condense in the tails of galaxies moving at high speeds soon after their central passage. In principle, many clumps could decelerate after ejection and end up depositing over time at $r=0$ if they lived for long enough, but it can be noted that this never happens. The time histogram has two dominant peaks, which correspond to the moments just after the central passage of the two most massive galaxies in the simulation. Those peaks go away in a timescale of a few hundreds of Myr, which is the lifetime of the clumps (most live between 100 and 300 Myr). The median clump velocity is 583 km/s -- by comparison, the velocity dispersion of the cluster (measured as the root mean square of the velocity distribution of its dark matter within the virial radius) is 959 km/s, evidencing that clumps are relatively decelerated objects. Indeed, below the median value there is a concentration of clumps which have decelerated and lurk around at low speeds before completely dissolving, and above this value there is a population of clumps which have just been ejected, and which are moving at speeds close to that of their progenitor galaxies ($\gtrsim$ 1000 km/s). Referring to those objects as ``high velocity clouds'' \citep[as in][]{2017ApJ...843..134S} is usually appropriate (only 7\% of the clumps move at less than 100 km/s). The passive scalar distribution shown in Figure \ref{fig:clumpskinematics} is very wide, showing that clumps can be found at varying degrees of mixing with the ICM.

\begin{figure}
 \includegraphics[width=\columnwidth]{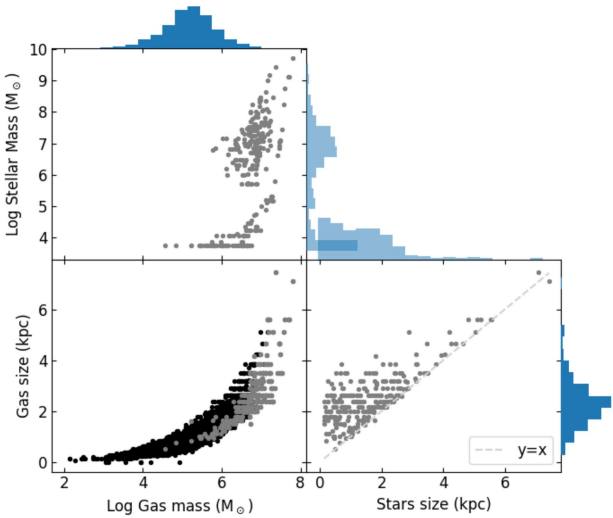}
 \caption{The masses and sizes of the clumps. Grey points represent clumps containing star particles, while black ones those which do not.}
 \label{fig:clumpsproperties}
\end{figure}

We define the size of a clump as the largest distance between its centre and the self-shielded cells in its bounding sphere. In the same manner, we also compute the size of its stellar content, defined as the largest such distance considering the positions of the stars within the sphere instead. The results are in Figure \ref{fig:clumpsproperties}, along with the masses of the clumps. Only a tiny minority of the clumps in the simulation contain star particles (3\%), and they are at the high end of the mass distribution. The extent of the stars in such clumps is smaller than that of the gas, indicating that the stars are formed at the centre of each clump and remain dynamically cold. Considering the histograms in this figure, we find that the typical gas mass of a clump is $1.6 \times 10^{5}$ M$_\odot$, with most clumps ranging from $2.7 \times 10^{3}$ M$_\odot$ to $4.6 \times 10^{6}$ M$_\odot$. The typical size of a clump is $2.2$ kpc, and most clumps range from $0.3$ kpc to $4.1$ kpc. The latter value seems high, but we note that the largest clumps tend to simply be irregular and elongated. Among the clumps which have stars, the typical stellar mass is $10^{7}$ M$_\odot$, and the typical gas mass is $7 \times 10^{6}$ M$_\odot$. A second, isolated peak can be seen in the stellar mass distribution; it is associated with a single relatively large mass of gas which was ejected from one of the galaxies after it entered the cluster in an almost radial orbit and crossed its central region at high speed.

\subsection{Galaxies}

The position of a galaxy in a snapshot is defined as the centre of mass of the star particles originally present in its initial conditions. Similarly to what we did with the clumps, we start our analysis with a sphere centred in each galaxy, but this time with a radius fixed at 10 times its radial scale length $R_d$. The non-self shielded cells in this sphere are discarded. Defining $\overrightarrow{v_{\mathrm{gal}}}$ as the average velocity of the original star particles, and $\sigma_{\mathrm{gal}}$ as the standard deviation of the velocities of these particles along the direction of $\overrightarrow{v_{\mathrm{gal}}}$, we also discard the gas cells and star particles for which 
\begin{equation}
\frac{|\overrightarrow{v}\cdot \overrightarrow{v_{\mathrm{gal}}}|}{||\overrightarrow{v_{\mathrm{gal}}}||} > 5 \sigma_{\mathrm{gal}}.
\end{equation}
This is a simple filter to discard any cells and stars in the sphere which are not moving along with the disk of the galaxy, and hence are not a part of it, e.g. those which belong to ejected clumps. 

\subsubsection{Gas contamination}

\begin{figure}
 \includegraphics[width=\columnwidth]{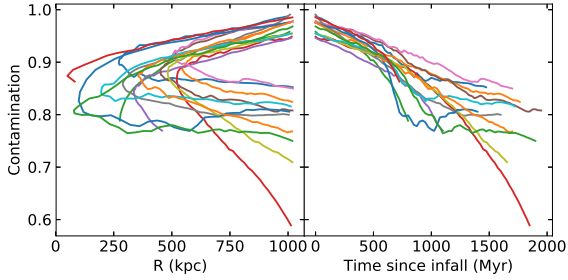}
 \caption{Change in gas composition in the disks of the galaxies as they cross the cluster. Similarly to Figure \ref{fig:clumpskinematics}, the passive scalar value shown is a mass weighted average. A value of 1 means that there is no contamination, and lower values represent mixing with the ICM.}
 \label{fig:galaxies}
\end{figure}

Using the gas cells remaining after this procedure, we extract the contamination of ICM gas within the disk of the galaxies as they cross the cluster, as shown in Figure \ref{fig:galaxies}. Some of the curves in this plot end before a complete crossing because the smallest galaxies are completely stripped. It can be seen that, after a crossing of the cluster, typically around 20\% of the neutral gas still in any given galaxy is actually ICM gas that advected there -- for reference, the median gas mass in the galaxies which have not been completely stripped is $39$\% of the original gas mass. In the most extreme case, the contamination reaches more than 40\%, but as we will discuss in Section \ref{sec:convergence} using a convergence test, we cannot rule out that this particular case is spurious. Ignoring any metal deposit from stars over time, we can extract from these values an upper bound for the predicted decrease in metallicity of the gaseous disks after one crossing of the cluster. We start by assuming the following typical metallicities: Z$_\mathrm{ICM} = 0.3$Z$_\odot$ \citep[which is consistent with observational data, e.g.][]{2001ApJ...551..153D} and Z$_\mathrm{ISM}$ = Z$_\odot \approx 0.02$. From that we obtain a typical decrease of 14\% in Z$_\mathrm{ISM}$, from 0.02 to $\sim$0.017.

\subsection{Diffuse light}

We compute the luminosities of the star particles formed into the clumps using the magnitude tables generated with \textsc{CMD} \citep[a web interface for the stellar evolution model described in][]{2017ApJ...835...77M} that are currently included in the \textsc{pynbody} package\footnote{See \url{http://stev.oapd.inaf.it/cgi-bin/cmd} and \url{https://pynbody.github.io/}}. These tables contain absolute magnitudes per solar mass in the UBVRIJHK system as a function of stellar age (assuming instantaneous star formation) and metallicity. We assume for simplicity a constant metallicity of $1$ Z$_\odot$ in our luminosity calculation, which is reasonable because the scatter in the stellar mass-metallicity relation is large \citep{2004ApJ...613..898T}, while 1 Z$_\odot$ is a common value. The magnitudes are then interpolated from the magnitudes in the tables using the ages of the star particles. 

\begin{figure}
 \includegraphics[width=\columnwidth]{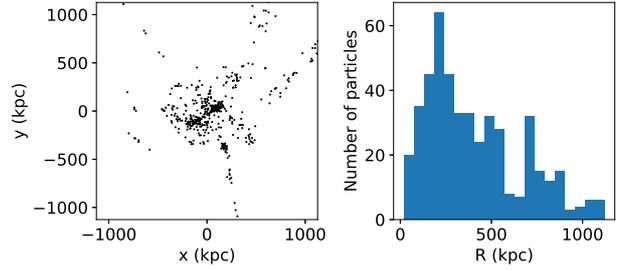}
 \caption{All the stars formed in the clumps, which live in the ICM unattached to any galaxy, as seen at the end of the simulation. They form a cloud with radius equal to $\sim500$ kpc at the centre of the cluster.}
 \label{fig:diffuse}
\end{figure}

The stars formed in those most massive clumps in Figure \ref{fig:clumpsproperties} continue to exist even after the clumps mix with the ICM. In Figure \ref{fig:diffuse} we show the spatial distribution of all such stars at the end of the simulation ($z = 0$). It can be noted that there is a central concentration of these stars in the cluster, for $r$ < 500 kpc. This happens because many of the stars have had time at this point to decelerate and fall back to the centre, where they interact with each other and form a cloud. The combined $M_\mathrm{V}$ of all the free particles is -10.8, while that of the particles within the inner 500 kpc of the cluster is -10.12. For comparison, our smallest galaxy ($M_{200c} = 1.1\times10^{11}$ M$_\odot$) has, before entering the cluster, $M_\mathrm{V} = -17.7$. Clearly the luminosity of those free stars is very low, leading us to conclude that the stars associated with the clumps are an insignificant component of the diffuse light of a galaxy cluster.

\subsection{Numerical convergence} \label{sec:convergence}

\begin{figure}
 \includegraphics[width=\columnwidth]{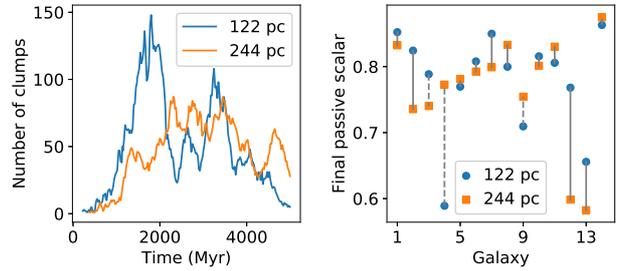}
 \caption{Convergence test. Left: number of clumps as a function of time for the two different resolutions. Right: comparison of the mass weighted average passive scalar in each galaxy for the two cases after one crossing of the cluster. Dashed lines indicate that in at least one of the simulations the galaxy was completely stripped before crossing.}
 \label{fig:convergence}
\end{figure}

In order to assert the numerical robustness of our results, we ran our idealised simulation again with everything unchanged except for the maximum level of refinement, which was reduced by one, resulting in a maximum resolution of 244 pc (compared to the 122 pc in the main run). In principle, a higher resolution implies in less numerical diffusion, better capturing of mixing and in the possibility of resolving smaller structures. In Figure \ref{fig:convergence} we compare the number of clumps in the cluster as a function of time for the two runs. What we observe is that in the high-resolution run, the condensation of clumps in the tails of the galaxies is visibly more intense, with more small clumps being generated. But at the same time, the mixing suppression dominates over numerical diffusion in the low resolution run, which gives rise to the intervals in Figure \ref{fig:convergence} where there are more clumps in it. Despite of these differences, throughout most of the simulation time the number of clumps is not too different between the two runs.

Due to the high speed of our galaxies, (> 1000 km/s), it could also be that the contamination of ICM gas within their gaseous disks we report is partly due to numerical diffusion. So we also compare the main simulation with the one at lower resolution to quantify this, as also shown in Figure \ref{fig:convergence}. It can be noted that our finding that a 20\% contamination happens after one crossing of the cluster is not resolution dependent, which implies that the accretion of ICM gas within the gaseous disks of the galaxies must be due to the cooling of ICM gas at the ISM/ICM interface. It should also be pointed that the most extreme case of contamination in the high-resolution run, where 40\% of the gas in the disk comes from the ICM, is much less pronounced in the lower resolution run, making this particular data point inconclusive.

\section{Discussion and summary} \label{sec:discussion}

We should point out that all the clump properties we obtain may have some uncertainty, given that at the clump scale processes beyond the scope of our simulations may play a crucial role in real galaxy clusters. For instance, previous simulations of cloud--shock interactions have demonstrated that the presence of magnetic fields \citep{2015MNRAS.449....2M} or heat conduction \citep{2007A&A...472..141V,2017MNRAS.470..114A} both significantly stabilise the cloud against disruption, while both effects are absent in our simulations. Moreover, it's possible that higher spacial resolutions would significantly change both the dynamics and the internal structure of the clumps. According to \citet{1994ApJ...420..213K}, a resolution of $\sim$100 cells per cloud radius is necessary for a fully converged solution in a cloud--shock simulation; for a typical clump with radius $\sim$1 kpc, this would imply in a 10 pc resolution, an order of magnitude smaller than the one we have managed to reach (122 pc), despite our resolution being, for current standards, outstanding for a simulation including an entire galaxy cluster. On the other hand, recent results by \citet{2018MNRAS.tmp.1177T} show that, for a cloud--shock scenario modelled after the cold clouds present in the Virgo cluster, no significant difference is found in the cloud dynamics for simulations with resolutions of 62.5 pc and 31.25 pc, which suggests that our simulation should not be too far from a converged solution. With those caveats pointed out, we believe our simulation setup at least demonstrates that ram pressure stripping is a viable mechanism for generating lone clumps of cold gas in the intracluster medium of galaxy clusters, although additional numerical work is needed to better constrain the internal structure and dynamics of clumps formed in this scenario.

By construction, the clumps we identify in our simulation are all self-shielded, and hence composed primarily of molecular Hydrogen. The amount of neutral Hydrogen in such clumps is expected to be non-negligible \citep[see e.g.][]{1983ApJ...264..546M}, making them presumably observable using radio observations of the 21 cm line. But finding a significant number of those clumps in a cluster is not guaranteed. Their number count fluctuates over time, since their lifetime is small compared to the age of the cluster, and since most of them are generated by large structures falling into it ($M_{200c} > 10^{12}$ M$_\odot$). In our model, that happened twice in a timespan of 5 Gyr. The peak number count of clumps in the whole volume of the cluster was 148, but there were also times late in the simulation when only 5 clumps were present. The overall average was of 47 clumps. We have made some videos available to illustrate the process of clump generation\footnote{\url{http://www.astro.iag.usp.br/~ruggiero/paperdata/2/}}.

Our clumps can be associated with objects already reported in the literature. The two clouds reported in the Virgo cluster in \citet{2010ApJ...725.2333K} have an H$_\mathrm{I}$ mass between $10^{7.28}$ and $10^{7.85}$ M$_\odot$, at the upper end of our mass distribution, which could be either random (given the small sample size) or an observational bias towards more massive objects. Another cloud reported in the Virgo cluster in \citet{2017ApJ...843..134S} has a stellar mass of $5.4\pm1.3 \times 10^4$ M$_\odot$, which is within our range of stellar masses. Its stellar age of $\sim 7-50$ Myr is relatively small compared to our clump lifetime. The clumps of molecular gas in the tail of NGC 4388 in Virgo reported in \citet{2015A&A...582A...6V} have mass between $7 \times 10^5$ and $2\times 10^6$ M$_\odot$, also within our distribution of gas masses. Similar objects found in groups of galaxies also agree with our results. In particular, \citet{2009A&A...507..723T} report compact, star forming objects with an age < 100 Myr in three groups, and \citet{2008AJ....135..319D} report similar objects in the Hickson Compact Group 100 with age < 200 Myr. Both of these are consistent with the clump lifetime we report. The stellar mass within such objects reported by \citet{2008AJ....135..319D} is $10^{4.3-6.5}$ M$_\odot$, which is within the distribution of stellar masses that we obtain; on the other hand, their H$_\mathrm{I}$ mass range, $10^{9.2-10.4}$ M$_\odot$, is above all of our clump masses. An important difference that should be noted between our clumps and those found in observed galaxy clusters is that the latter tend to be found preferentially at the periphery of their host clusters, while our clumps tend to be formed preferentially near the centre of the cluster ($r < R_{200c}/2$); this may mean that we are underestimating the ram pressure intensity at the cluster outskirts, possibly because of the assumption of hydrostatic equilibrium (which may lead to lower relative velocities between the galaxies and the ICM).

It should be emphasised that in this work we have considered the evolution of a relatively undisturbed cluster gradually accreting small satellites over time. But observationally, a systematic computer vision search for jellyfish morphologies \citep{2016MNRAS.455.2994M} has found that such systems seem to be preferably found in disturbed environments, such as cluster mergers. In such scenarios, the relative velocities between the galaxies and the surrounding medium is higher, possibly leading to higher ram pressure intensities, and hence possibly different results from what we find here -- for instance, the clump generation could be triggered much earlier, already at the periphery of the clusters involved, reducing the preference we find for the inner part of the cluster.

The lifetime of the clumps we find in our simulation, coupled with their fast deceleration after ejection from the galaxy, is consistent with the finding from \citet{2008MNRAS.388..465R} that the gas lost by the galaxies tends to be deposited not far from where it was lost. One interesting inference we can make from this and from the clump dynamics we show in Figure \ref{fig:clumpskinematics} is that the molecular gas resulting from a ram pressure stripping event is not expected to be able to feed a galaxy cluster's central AGN, since not a single clump manages to reach r = 0. The only way we can picture the infalling galaxies influencing the cluster's AGN is if one of them fell radially and hit the black hole directly, something that to our knowledge has never been reported.

One could wonder if the clump lifetime we report (100--300 Myr) could be related to a Kelvin-Helmholtz (KH) instability disruption timescale. Following \citet{1959flme.book.....L}, this timescale can be written in terms of the ICM density $\rho_{\mathrm{ICM}}$, the clump density $\rho_\mathrm{clump}$, the relative velocity between the two $v$ and the instability wavelength $\lambda$ (i.e. the eddy size) as
\begin{equation}
T_{\mathrm{KH}} \sim \frac{\rho_{\mathrm{ICM}}+\rho_\mathrm{clump}}{\sqrt{\rho_{\mathrm{ICM}}\rho_\mathrm{clump}}} \frac{\lambda}{v}.
\end{equation}
An upper bound for this timescale can be obtained considering $\lambda$ equal to the typical clump size (2.2 kpc). Considering as typical values $v = 500$ km/s (see Figure \ref{fig:clumpskinematics}), $\rho_{\mathrm{ICM}} = 2\times10^{-28}$ g/cm$^3$ (which is the density at r = 500 kpc, considering Eq. \ref{eq:betaprofile}) and $\rho_{\mathrm{clump}} = 5 \times10^{-27}$ g/cm$^3$ (which is the self-shielding threshold we adopt), we obtain
\begin{equation}
T_{\mathrm{KH}} \lesssim 20 \mathrm{Myr},
\end{equation}
an order of magnitude smaller than our clump lifetime. This difference can be attributed to two factors. First, and most importantly, the mass of the clumps is high enough for their self-gravity to counter-effect the KH instability. As shown in \citet{1993ApJ...407..588M} \citep[see also][]{1956MNRAS.116..351B}, the critical mass for a pressure-bound isothermal sphere to be gravitationally unstable is
\begin{equation}
M_{G} = 1.18 G^{-3/2} \left(\frac{k_B T_\mathrm{clump}}{\mu m_H}\right)^{3/2} \frac{\rho_\mathrm{ICM}^{3/2}}{\rho_\mathrm{clump}^2}.
\end{equation}
Considering as typical values $T_\mathrm{clump} = 8000$ K and $\mu = 1.2$, this results in $M_G \sim 10^6$ M$_\odot$, a number within the upper extreme of our clump mass distribution. The fact that this number is not too much larger than our clump masses implies that self-gravity is relevant to their dynamics. At the same time, the clumps also gain molecular mass over time due to the condensation of ICM gas at their interfaces, as evidenced by the passive scalar distribution shown in Figure \ref{fig:clumpskinematics}, which shows that ICM gas is a significant component of those objects. This also causes the rate of gas loss due to KH instability to be suppressed.

Our result that few of our clumps (3\%) have stars is biased, since our smallest particle mass is $5.6 \times 10^{3}$ M$_\odot$. It's possible that smaller clumps do have stars, but in minute amounts not resolved by our resolution. In fact, \citet{2013ApJ...767L..29O} has managed to observe a single star in Virgo which belongs to a system analogous to the ones we describe here, a case that could not be covered by our numerical methodology. With that pointed out, the star formation inefficiency we obtain is consistent with both observations of ram pressure wakes \citep{2015A&A...582A...6V,2017ApJ...839..114J} and with previous simulations of such systems \citep{2012MNRAS.422.1609T}, and it is the reason why the star formation in the tails ends up being a negligible source of intracluster light.

Concerning the contamination of the disks of the galaxies with ICM gas, we have reason to believe that the process should undergo similarly in galaxy clusters or groups across a range of masses. This is because the final contamination is roughly the same for galaxies which enter our cluster with a range of different impact parameters (as robustly evidenced by our convergence test, Figure \ref{fig:convergence}), and thus encounter a range of different surrounding densities, temperatures and relative velocities. Despite those differences, the galaxies are contaminated over time at a rather similar rate (as seen in Figure \ref{fig:galaxies}). This suggests that what primarily governs the contamination are not the particularities of the surrounding gas, but the time since the infall of the galaxy instead.

The summary of this paper is as follows. Using a high-resolution hydrodynamic simulation of a galaxy cluster, we found that the gas stripped from infalling gas-rich galaxies gives rise to a population of clumps of molecular gas in the ICM, which live for long enough to not seem obviously associated to any galaxy in the cluster. The exchange of gas also goes the other way around: an infalling galaxy advects gas from the ICM as it moves, and after one crossing of the cluster typically around 20\% of the gas in its disk is actually ICM gas that ended there. We also find that the stars that form in the ram pressure tails of the galaxies are a completely negligible source of intracluster light.

\begin{figure*}
 \includegraphics[width=\textwidth]{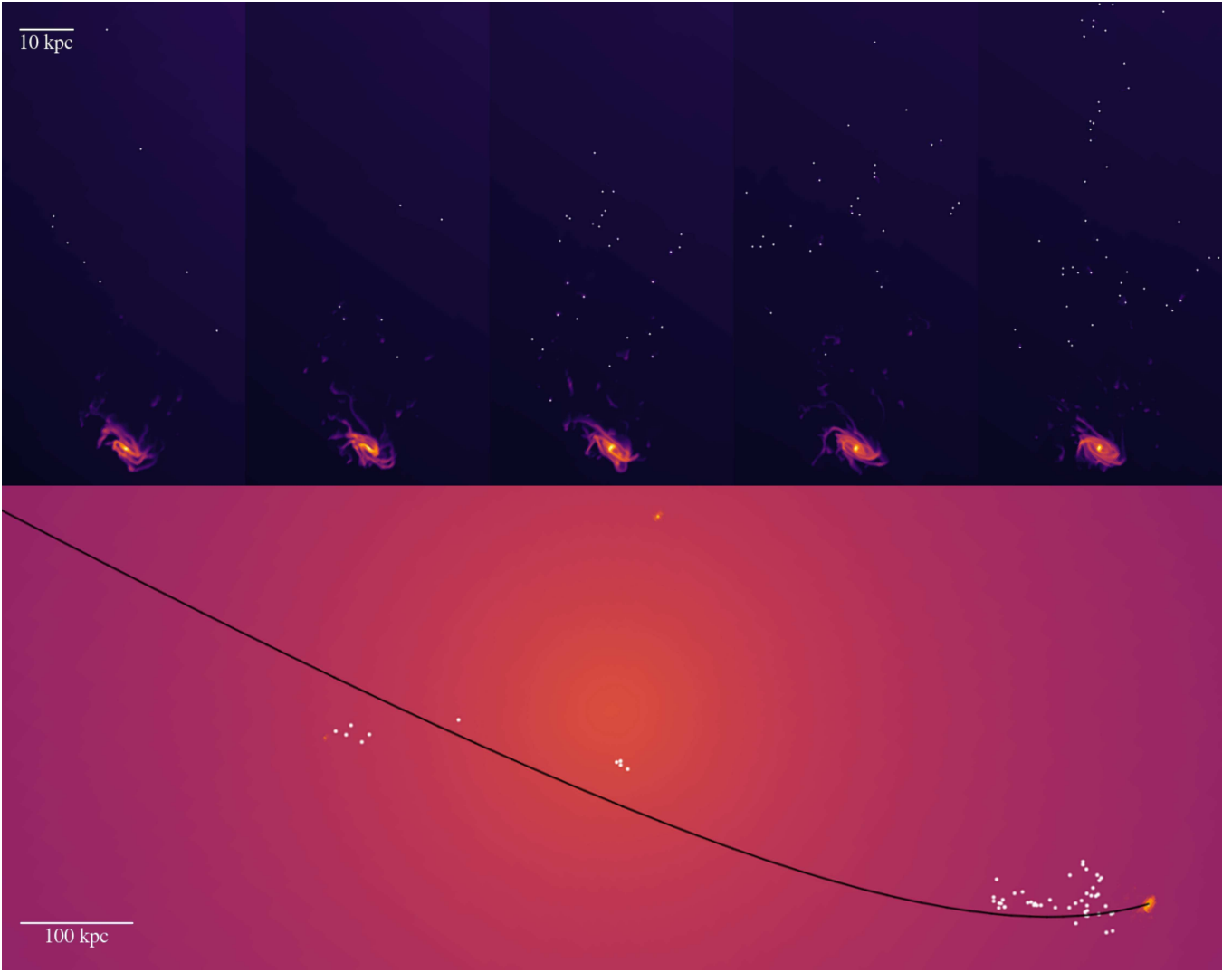}
 \caption{Top of the image: a time series showing the ejection of clumps by one of our galaxies after its central passage. The time difference between each frame and the next is 100 Myr. Bottom image: a zoomed out version of the last frame showing the galaxy and the cluster as a whole. The clumps further to the left of the image come from other galaxies. A video showing the clumps moving within the cluster over time can be watched at \url{http://www.astro.iag.usp.br/~ruggiero/paperdata/2/}.}
 \label{fig:panel}
\end{figure*}

\section*{Acknowledgments}

The research project of which this work is a part is funded by the S\~ao Paulo Research Foundation, FAPESP (grants 15/13141-7 and 16/19586-3). GBLN thanks CNPq for partial financial support. The simulations presented here were run at the Swiss National Supercomputing Center (CSCS), specifically using the Piz Daint supercomputer. Our data analysis has been largely done using the Hydra cluster at the University of Zurich, and it has extensively used \textsc{yt} \citep{2011ApJS..192....9T}. We thank Daisuke Nagai and Tom Abel for the suggestions which helped shape this paper, and the anonymous reviewer for the constructive feedback.

\bibliographystyle{mnras}
\bibliography{references}

\bsp	
\label{lastpage}
\end{document}